\documentstyle[12pt,twoside]{article}
\def\greaterthansquiggle{\raise.3ex\hbox{$>$\kern-.75em\lower1ex\hbox{$\sim$}}}
\def\lessthansquiggle{\raise.3ex\hbox{$<$\kern-.75em\lower1ex\hbox{$\sim$}}}
\newcommand{\beq}{\begin{equation}}
\newcommand{\eeq}{\end{equation}}
\newcommand{\beqa}{\begin{eqnarray}}
\newcommand{\eeqa}{\end{eqnarray}}
\newcommand{\beqan}{\begin{eqnarray*}}
\newcommand{\eeqan}{\end{eqnarray*}}
\newcommand{\ba}{\begin{array}}
\newcommand{\ea}{\end{array}}

\newcommand{\T}{{\cal T}}

\def\nz{\ifmmode {I\hskip -3pt N} \else {\hbox {$I\hskip -3pt N$}}\fi}
\def\zz{\ifmmode {Z\hskip -4.8pt Z} \else
       {\hbox {$Z\hskip -4.8pt Z$}}\fi}
\def\qz{\ifmmode {Q\hskip -5.0pt\vrule height6.0pt depth 0pt
       \hskip 6pt} \else {\hbox
       {$Q\hskip -5.0pt\vrule height6.0pt depth 0pt\hskip 6pt$}}\fi}
\def\rz{\ifmmode {I\hskip -3pt R} \else {\hbox {$I\hskip -3pt R$}}\fi}
\def\cz{\ifmmode {C\hskip -4.8pt\vrule height5.8pt\hskip 6.3pt} \else
       {\hbox {$C\hskip -4.8pt\vrule height5.8pt\hskip 6.3pt$}}\fi}

\def\au{{\setbox0=\hbox{\lower1.36775ex%
\hbox{''}\kern-.05em}\dp0=.36775ex\hskip0pt\box0}}
\def\ao{{}\kern-.10em\hbox{``}}

\normalsize
\begin{document}
\bibliographystyle{plain}

\begin{titlepage}
\begin{flushright} UWThPh-2004-9\\

\today
\end{flushright}
\vspace*{2.2cm}
\begin{center}
{\Large \bf  Fluctuations of quantum lattice systems under Boltzmann evolution}\\[30pt]

Heide Narnhofer  $^\ast $\\ [10pt] {\small\it}
Faculty for Physics  \\ Vienna University\\

\vfill \vspace{0.4cm}

\begin{abstract}For quantum lattice systems a Boltzmann type time evolution arises according to results of Hugenholtz in the limit of N-scaled time evolution together with an interaction scaled as $N^{-1/2}$. According to Illner -Neunzert this passage to an irreversible dynamic can only happen when the initial state satisfies stronger assumptions than the evolved state. We show that the initial state allows to control space clustering as it is the minimal requirement for the construction of a fluctuation algebra in the sense of Verbeure et al.,  whereas in the evolved state fluctuations only are defined in an appropriate scaling local with respect to the time evolution but not on a mesoscopic level.


\hspace{1.9cm}

\end{abstract}
\end{center}

\vfill {\footnotesize}

$^\ast$ {E--mail address: narnh@ap.univie.ac.at}
\end{titlepage}

\section{Introduction}
The fact that in thermodynamics time evolution is irreversible is undoubted. Also the fact that we should understand thermodynamics as resulting from n-particle physics and its thermodynamic limit is taken for granted. An important and extremely successful step is provided by the Boltzmann equation. It combines particle physics and its scattering mechanism expressed by the scattering cross section and statistics expressed by the density function. Initially it was given as ansatz based on plausibility \cite{B}, but it is possible to deduce it rigorously as a limit where the number of particles goes to infinity though in a way that the particle density is small enough so that the scattering process remains an individual process just between two particles \cite{L}. The mathematical control is still limited to essentially some finite number of collision processes and this can be achieved by various assumptions on the initial state. Still there are attempts to vary and reduce these assumptions \cite{P}. But in applications the Boltzmann equation is so successful to describe the actually observed behaviour that there is good reason to believe that the limitations in the proof are given by mathematical difficulties and not by conceptual errors.
Nevertheless from the very beginning there were controversial considerations how the reversible time evolution of n-particle physics can transform into the irreversible time evolution given by the Boltzmann equation. This irreversibility is manifested in the increase of entropy. However the monotonic behaviour of some function with respect to time is not jet an appropriate characterisation of irreversiblity, already for one particle the dilation $x(t)p(t)=x(0)p(0)+p^2(0)t/2$ increases in time. In \cite{IN} this question of irreversibility was investigated. On one hand the authors clarified in which sense irreversibility has to be understood in order to be relevant for thermodynamics. Next they inspected the proof for the derivation of the Boltzmann equation and observed, that in the proof the demands on the convergence of the density function for the starting state is stronger than for the evolved states. Finally they showed that this is not only an inconvenience due to the fact that we did not find the appropriate mathematical arguments but it is essential: only if initial state and evolved state are defined in different topologies the time evolution can undergo such an essential qualitative change from reversibility to irreversibility.

Boltzmann equation is an evolution equation in the framework of classical physics. \cite{H} considered a corresponding approach in the frame work of quantum mechanics. He studied Fermions on an infinite lattice where we have the advantage that free evolution and also time evolution with interaction is well defined as automorphisms on the corresponding $C^*$ algebra. In contrast to the classical assumptions we do not consider a  limit corresponding to an increasing particle number. The thermodynamic limit is already given by starting with the $C^*$ algebra. Therefore low density is no suitable concept. But the idea that the scattering process is an individual process between pairs can be adopted. Also in the classical situation we can think of two time scales, one the time in which the scattering process is finished, which is given by  the size of the cross section which in the approach tends to a point, and the time in which the particles change their location.
In a similar way \cite{H} considered two time scales: the time scale necessary for the scattering process being expressed only by the twopoint function,
which according to the velocity asks for a weak interaction, and a time scale so that in spite of the weakness of the interaction the effects can summarize. With the appropriate scaling the twopoint function, which in this setting is just a function of momentum and not of space, evolves by a differential equation very similar to the Boltzmann equation. Especially this evolution has the same advantage that the entropy density increases and that the fixed point of the evolution are equilibrium states of the free unperturbed evolution. The essential steps will be given in section (2).

We are interested how in this approach the result of \cite{IN} can be observed. We will notice that the difference in topology is given in the way in which the twopoint function has to be interpreted. But in addition we can use an additional concept to see in which sense information gets lost: a state cannot only be interpreted as a state over the quasilocal $C^*$algebra. It can be enlarged to define a state on mesoscopic observables, the fluctuation algebra introduced in \cite{V}. Its definition will be given in section (3). This is not always possible but asks for strong spacial clustering properties that however can be provided in appropriate states as for example equilibrium states \cite{M}. This fluctuation algebra exists for the initial state satisfying the demands in \cite{H}. In section (4) we will concentrate on this fluctuation algebra, where we have the additional option to define it in various ways corresponding to the scaling in the evaluation of the Boltzmann equation. We will observe how the control on long range spacial correlations though unobservable on the quasilocal level but necessary for the construction of the fluctuation algebra gets lost in the cause of time.
\section{ The approach of Hugenholtz}
We repeat the definitions and relevant results of \cite{H} as they are the basis and ask only for minor generalisations for the fluctuation algebra.

We consider the Fermi-algebra on a lattice, built by creation and annihilation operators $a(x), a(y)^*$ satisfying the anticommutation relations \beq [a(x),a(y)]=0,\quad [a(x),a(y)^*]=\delta _{xy} \eeq
and the corresponding operators $a(f)=\sum _x f(x)a(x)$ satisfying $||a(f)||=\sum _x |f(x)|^2.$ In addition we will calculate with their Fourier transform $a_p$ where we have to keep in mind that these are not operators but unbounded forms well defined in appropriate representations. We start from the very beginning with a quasifree state that is translation invariant given by the twopoint function
\beq \omega (a(f)^*a(g))=\sum _{x,y} w(x-y)\bar{f(x)}g(y)).\eeq
In \cite{H} Hugenholtz started with more generality but showed that with appropriate assumptions in the limit the truncated functions vanish.
In the following we will start from the very beginning with such states and work in momentum space. We consider a quasifree evolution
\beq \tau_0(t)a(f) =a(e^{iht}f)=\int _{-\pi }^{\pi} dp e^{i\epsilon(p)}f(p)a_p \eeq
and an interaction given in Fourier space by \beq V=\int dk dl dm dn [v(k-n)-v(k-m)]\delta (k+l-m-n)a_k^*a_l^*a_ma_n.\eeq
For this interaction it is well known that its expression in x-space allows to show by perturbation theory that also in combination with the quasifree evolution it defines an automorphism group on the $C^*$ algebra. Summations and integrations can be understood in various dimensions without effecting the final result.

The Boltzmann evolution is obtained by considering the evolution $\tau (t)$ implemented by \beq H_{\lambda ,N }= \epsilon(p)a_p^*a_p+ N^{-1/2}\lambda V\eeq
and considering \beq \omega _t(A)=\lim _{N\rightarrow \infty }\omega(\tau (tN)A)\eeq
The interaction is treated on a perturbative level expanding in $\lambda $ with multicommutators
\beq \tau ^0_{-Nt}\tau ^{\lambda } _{tN} A=\sum _n (i\lambda )^n\int _0 ^{tN}dt_1 \int _0^{t_1}..[\tau ^0 _{t_1}(N^{-1/2}V), [\tau ^0 _{t_2}(N^{-1/2}V),...A]..\eeq
To first order for our starting quasifree state we get a sum over expectation values of operators such that the expectation value  splits into products of the expectation value of  pairs of creation and annilhilation operators. Now the invariance under space translations together with the symmetry relations resulting from the commutators implies that to first order of $\lambda $ together with the scaling in $N$ this term vanishes, and the same remains true for all contributions of odd order in $\lambda .$ For even contributions again we concentrate on the twopoint contributions that contain a factor $e^{i\epsilon(m)t_r-i\epsilon (l)t_s}$ or only $e^{i\epsilon (m)t_r}$. Integration over $t_r$ let these terms vanish in the limit $N\rightarrow \infty$ in combination with the scaling of the interaction. Exceptions appear if for the integration with respect to $t_r$ and $t_s$ the time dependence in the exponent is opposite to one another so that we can split into $dt_r-dt_s$ which gives a finite integral due to asymptotic abelianess and the assumption on the clustering of the state  and the integral over $dt_r+dt_s$ which is of order $N$ and multiplied with the scaling of the interaction also gives a finite term and therefore contributes to the change of the state. This coincidence happens when the double commutant of the interaction acts on the same annihilation respectively creation operator. We refer to (4.9) in \cite{H}. The various terms can be represented by diagrams that show that  the number of contributing  terms is sufficiently small to guarantee convergence of the perturbation theory. In addition only those terms contribute where the contributions of creation and annihilation operators defining the operator $A$ in (7) enter in the same combination  as in the starting state, though with different weight corresponding to the interaction. Therefore the limit state is again quasifree but with a different twopoint function. All details are given in \cite{H}.

 With
\beq \omega _t(a(f)^* a(g)=\int dp w(p,t)\bar{f(p)}g(p)\eeq
and concentrating on $t\rightarrow 0$ \cite{H} obtained the differential equation and extended it to $t>0$
\beq \frac{d}{dt} w( p,t)= \pi \int dkdldm|\langle kl|V|mp\rangle |^2 \delta (k+l-m-p)\delta (\epsilon _k+\epsilon_l -\epsilon _m -\epsilon _p) \eeq
$$(w(k,t)w(l,t)(1-w(m,t))(1-w(p,t))-w(p,t)w(m,t)(1-w(k,t))(1-w(l,t))).$$
Notice that this extension from $t=0$ to arbitrary $t$ has to be justified because according to the observation of \cite{IN} the state at a later time will not satisfy the same assumption as those that are needed for $t=0.$
\section{The fluctuation algebra}
In \cite{V} the fluctuation algebra was introduced as an additional algebra that can be associated to the quasilocal algebra in states that are space translation invariant and satisfy additional appropriate properties. We consider with space translation $\alpha _x A=A_x$
\beq \lim _{K\rightarrow \infty }\omega (e^{iK^{-1/2}\sum _{x=1}^K \alpha _x A-iK^{1/2}\omega (A)})=\bar{\omega }(W(A))\eeq
where $W(A)$ can be interpreted as a unitary operator  depending on the local operator $A$ with expectation value in the state $\bar{\omega }$. Restriction to local operators is useful to insure that the limit can exist but can be extended to quasilocal operators. In the same spirit we can extend the construction to $\bar{\omega }(W(A)W(B))$ and observe that the operators satisfy \beq W(A)W(B)=e^{\omega ([A,B])}W(B)W(A)\eeq
so that the unitary operators $W(A)$ form a bosonic Weyl-algebra. The state $\bar{\omega }$ can now be interpreted as a state on both the quasilocal algebra and the corresponding Weyl-algebra, which is itself state dependent. With
\beq  \bar{\omega }(W(A)BW(-A))= \omega (B) \eeq
 whereas
 \beq \bar{\omega }(W(A)W(B)W(-A))=e^{\omega ([A,B])}\bar{\omega }(W(B)),\eeq obviously the same state on the quasilocal algebra can give rise to different states on the fluctuation algebra, therefore some specification is needed, and this specification will give the difference between the initial state for the Boltzmann evolution and the evolved state.

 The additional properties guaranteing the existence of the limit in (10) together with the possibility to interpret is as the state over a fluctuation algebra ask for good control on the spacial decay of the state on the quasilocal algebra. That they can be satisfied can easily be seen on the example of a product state $\omega (A_xB_y)=\omega (A_x)\omega (B_x)$ where with $\omega (A_x)=0$
 \beq \bar{\omega }(W(A))=\lim _{K\rightarrow \infty }(1-\omega (A_x^2)/K)^K=e^{-\omega (A_x^2)}.\eeq
 In more generality it was proven in \cite{M} that the limit (10) can be controlled for KMS-states constructed from appropriate Hamiltonians. We are mainly interested whether there are additional conditions for quasifree states for the control of (10) and they can be seen best in momentum space: with
 \beq W (a^*(f) a(f)) = \lim _{K \rightarrow \infty }e^{i\sum _x e^{ix(p-q)}K^{-1/2}a^*_p a_q \bar{f(p)} f(q)dpdq} \eeq
  we observe that $w(p)$ has to be interpreted as a distribution and not just as an $L_1$ function in the same way as it had to guarantee that contributions of different momenta vanish in the intergration over time in (4.9) of \cite{H} and not only due to the smearing effect of the state.
  \section{The limit for the fluctuations}
  Evaluating the fluctuation algebra and its expectation value corresponding to the state on the quasilocal algebra we concentrate on

\beq \frac{d}{d\alpha }\bar{\omega }(W(\alpha A) |_{\alpha =0}=\lim _{K\rightarrow \infty } \omega (\sum _0^K K^{-1/2}\sigma _x A)-K^{1/2}\omega (A) \eeq
and
\beq \frac{d^2}{d\alpha ^2}\bar{\omega }(W(\alpha A) |_{\alpha =0}=\lim _{K\rightarrow \infty } \omega (\sum _0^K K^{-1}\sigma _x A)\sum _0^K\sigma _y A) -\omega (A)^2\eeq
and their evolution under the time evolution leading to the Boltzmann evolution, i.e.\beq  \lim _{N\rightarrow \infty }\lim _{K\rightarrow \infty } \omega (\tau _{-Nt}^0 \tau _{Nt}^{\lambda}K^{-1/2}\sum _0^K \sigma _x A)-K^{1/2}  \omega _t(A)\eeq
with
\beq \lim _{N\rightarrow \infty } \omega (\tau _{-Nt}^0 \tau _{Nt}^{\lambda}A) =\omega _t(A)\eeq
respectively

\beq \lim _{K\rightarrow \infty }\lim _{N\rightarrow \infty } \omega (\tau _{-Nt}^0 \tau _{Nt}^{\lambda}K^{-1/2} \sum_0^K \sigma _x A)-K^{1/2}\omega _t(A) .\eeq
and especially

\beq \lim _{N\rightarrow \infty } \omega (\tau _{-Nt}^0 \tau _{Nt}^{\lambda}N^{-1/2} \sum_0^N \sigma _x A)-N^{1/2}\omega _t(A) .\eeq
We follow the expansion of Hugenholtz  expressed as an expansion in $\lambda.$ A typical $A$ including the translation in space has the form
\beq \sum _{z_1,..u_1, x}\int dk_1..dl_1a^*(k_1)..a^*(k_r)..a(l_1)exp(ik_1z_1+..-il_1u_1)exp((ik_1+---il_1)x) \bar{f(k_1)}..g(l_1).\eeq
We concentrate mostly on $r=1$, i.e. on a quadratic $A$. Inserting this into (4.3) of \cite{H} and starting from the very beginning with a quasifree state invariant under the unperturbed time evolution
from space invariance of the state we keep the $\delta$-distribution for $(k_1-l_1)$ and similarily for all other terms expressing the contributions coming from the interaction in the expansion. Independent in which order we take the limit in (18),(19) and (20) it tends to $0,$ since in the approach of \cite{H} the fact, that the limit state is space translation invariant, remains unchanged.

 However for the existence of the fluctuation algebra $W(\alpha A)$ is interpreted as a unitary operator  whose second derivative is well defined so that (17) has to remain bounded. Therefore we have to take into account that the combination of integration in $t_r$ and summation over $x$ gives rise to additional contributions similar as the combination of integration in $t_r$ and $t_s$ was responsible for the Boltzmann evolution on the quasilocal level. Whereas on the quasilocal level only contributions corresponding to
 \beq \int dt_r dt_s [V(t_r),[V(t_s) a^*(f)].. a\eeq
 had to be taken into account because only here the exponant in the $t_r$ and $t_s$ integration are properly related, now also with the appropriate scaling as it enters in (17)
 \beq K/N\sum _x^K\int dt_r dt_s [V(t_r), a^*(f)][V(t_s),a(g_x))]\eeq  contributes. Appropriate combinations of twopoint functions give
 \beq \sum ^K \int^N dt_r \int^N dt_s e^{i(\epsilon (l_r)-\epsilon (m_r))(t_r -t_s)}e^{i\epsilon (k_1)t_r}e^{-i\epsilon(k_2)t_s)} e^{i(k_1-k_2)x} .\eeq

 Summation over $x$ reduces to values in the integration over the momenta where $k_1-k_2 $ has to be of order $1/K.$ This effects the integration over $t_r$ and $t_s$ only mildly as long as $K<<N$. However it is leading to $e^{i\epsilon (k_1)(t_r -t_s)}$ up to negligible order if $K\geq N $ and is therefore of the same type as the contributions leading to the Boltzmann equation. Therefore the total integration and summation give additional terms of order $K/N.$

 If for the initial state the fluctuation algebra is well approximated already for $K$ and the interaction is weak compared to $K$, i.e. $N>>K$ then also in the scaled time intervall the fluctuations over the corresponding region remain regular. If the interaction is comparable  to $K$, $N=\gamma K$, then the fluctuations evaluated in a region comparable to the spreading of the particles remain finite but different from the value that is expected from the state on the quasilocal algebra, depending on $\gamma $ and therefore also on $t.$  If however we consider $K>>N$, which corresponds to the assumption that in the initial state the fluctuation algebra exists as limit in $K$ then we observe that in the evolved state the fluctuation algebra cannot be constructed.

It remains to control at which level the construction of the fluctuation algebra fails. Here we study \beq  \lim exp(i\sum _x^K K^{-1/2}\sigma _x A- iK^{1/2}\omega (A))=W(A)\eeq with selfadjoint $A$ and interpret the limiting operator as the exponential of a Bosefield  with Bosetype commutation relations with operators obtained in the same limit process. This holds for $t=0$ according to \cite{M} for our choice of $\omega .$ In addition this Bose field is in a gauge invariant quasifree state determined by the spacial correlation functions of the underlying quasilocal state. First we stay in the region where the time evolution is not scaled and  can be interpreted as a time evolution on the Bosefield. Let us choose the scaling of the interaction $N>>K$. With the scaling of the interaction the commutator with the Bosefields vanish and the interaction has no effect . Of interest is the area where $N$ and $K$ are of the same size. The Bose field is obtained by averaging over a region of size $N$, whereas the interaction acts everywhere and the free time evolution also shifts the Bosefield.  Therefore we have to consider the Bosefield to be a field on which space translations, now scaled by the factor $N^{-1}$ act. For finite times
\beq \tau _{-t}^0 \tau _t^{\lambda }W(G)= W(A)e^{i\int^t dt'\omega ([V(t'),A])}.\eeq
The commutator reduces to a c-number  and the interaction has the effect of a linear shift on the fluctuation algebra as long as $t$ remains small. If however we increase the time evolution to $N^{1/2}t$, then as in \cite{B} we have to take into account that $A$ is shifted outside of the previous region. The commutator with the time translated potential creates a finite contribution of products of operators with a distance of order $N^{1/2}$ in the exponential that are not controlled by the quasifree state on the quasilocal algebra.  The scaled space translation becomes effective  in addition to the time evolution and the Bosefield starts to change apart from a linear shift.

We have concentrated on the effect of the evolution on the fluctuation algebra in space dimension $1$ because here the scaling of the interaction and the scaling for the fluctuation algebra coincide. In higher dimensions the appropriate scaling for the fluctuation algebra has to be adjusted to $K^{ \nu/2}$. The scaling for the time evolution remains the same, because it is determined by the demand that the contribution in first order
of the perturbation in $\lambda $ has to vanish. This has the only consequence that the relation between time and the area, where spacial correlations fail to give rise to a fluctuation algebra changes.

There remains a last task: the evolution of the state on the quasilocal level is expressed in \cite{H} as a differential equation for the twopoint function. However the form of this equation is only evaluated for $t=0$ and extrapolated to $t$, though we have seen that $\omega _t$ does not satisfy the necessary demands to justify the limit when considered as starting point.  However we can use for the quasilocal state and time ordered integration
\beq \frac{d}{dt}\omega _t (G) =N^{1/2}\omega (V(t)\T \int dt'..[..( V(t'),G)]]).\eeq
Using that $\omega $ is quasifree and combining in the expansion in $\lambda $ all twopoint functions  apart from those of $V(t)$ the combinatoric keeps its form  and the differential equation for the twopoint function remains as for the initial state.

\section{Conclusion}
Following the strategy in \cite{H} we have extended the rate of convergence to operators that indicate the amount of spacial correlations as they are necessary for the construction of the fluctuation algebra. This fluctuation algebra allows to  compare the range of spacial correlations with the time scale leading to the Boltzmann evolution . Due to the scaling of the interaction these spacial correlations remain unchanged for finite times. But if we give the interaction enough time so that the scattering process between a pair of particles can become effective, which is the inverse time scaling of the scaling of the interaction, then already the spacial correlations change. This change is too small to be observed on the local level, but can be recognized if we pass to the mesoscopic observables of the fluctuation algebra. With increasing time the fluctuation algebra cannot be constructed any more and we loose information about mesoscopic observables. If however we consider the fluctuation algebra in a scaling that is small compared to the distance particles move then the averaging in the construction is sufficient and the fluctuation algebra survives. Interpreting this observation in the language of \cite{IN}, the two point function of the initial state is given as a continuous function in  distributional sense whereas for the  evolved state it only makes sense as an $L_{\infty }$ function and is therefore defined in a different, weaker topology. However the control on the possible scaling
together with the effect of averaging in space as it is  offered by studying the fluctuation algebra makes the change in the quality of the twopoint function more precise.

 We can compare the results with those in \cite{B}, \cite{BC} for a mean field Lindblad evolution as an example, where the time evolution violates irreversibility and nevertheless can be obtained in an appropriate setting as the limit of unitary time evolutions or also for a mean field automorphic time evolution. Here the
 evolutions created a change of the quasilocal state producing a map between fluctuation algebras with different commutation relations, however the construction of the fluctuation algebra was still possible though producing a slightly different state for the fluctuation algebra. Therefore spacial correlations preserved their character up to a small variation. Especially for a state over the quasilocal algebra that remained constant under  the mean field time evolution the state  for the fluctuation algebra the state changed in time though the fluctuation algebra survived. Here the long range behaviour of the interaction in its relation to the definition of the fluctuation algebra was responsible for the different behaviour. Now the state on the quasilocal level changes in an appropriate scaling. Depending how the fluctuation algebra can be defined with respect to this scaling its state changes or even the construction of the algebra fails. Spacial correlations over mesoscopic distances i. e. distances over which particles move in order to feel the effect of the interaction get out of control but only to a fine tuning that cannot be observed on the microscopic level or under the macroscopic level under a rough averaging.

\bibliographystyle{plain}

\begin{thebibliography}{99}
\bibitem{B} L. Boltzmann: Lectures on Gas Theory, Englich addition
\bibitem{L} O.E. Lanford III: On a Derivation of the Boltzmann Equation, in Nonequilibrium Phenomena I -The Boltzmann Equation, North Holland (1983)
\bibitem{P} R.Illner, M. Pulvirenti: Global Validity of the Boltzmann Equation for a For a Two- and Three Dimensional Rare Gas in Vacuum,Commun.Math. Phys. {\bf121},(1989)
\bibitem{IN} R. Illner, H. Neunzert: The Concept of Irreversibility in the Kinetic Theory of Gases, Transport Theory of Statistical Physics {\bf16/1} 89-112 (1987)
\bibitem{H} N.M. Hugenholtz: Derivation of the Boltzmann Equation for a Fermi Gas, Journ Stat. Physics, {\bf32/2} 231-254 (1983)
\bibitem{V} D. Goderis, A.Verbeure, P. Vets: Commun. Math. Phys. {\bf128}533549 (1990)
\bibitem{M} T. Matsui: Rev. Math. Phys. {\bf14}675, (2002)
\bibitem{B} F. Benatti, F. Carollo, R. Floreanini, H. Narnhofer: Quantum Fluctuations in Mesoscopic Systems J. Phys. A: Math. Theor {\bf50}423001 (2017)
\bibitem{BC} F. Benatti, F. Carollo, R. Floreanini, H. Narnhofer: Phys. Lett. A {\bf326} 187 (2014)






\end{thebibliography}

\end{document}